\documentclass[a4paper,11pt]{article}
\usepackage{jinstpub} 
\usepackage{lineno}

\title{Development of an X-ray polarimeter at the SOLEIL Synchrotron}

\author[a]{L. Manzanillas,}
\author[a]{J.M. Ablett,}
\author[a]{M. Choukroun,}
\author[a]{F.J. Iguaz,}
\author[a]{J.P. Rueff}
\affiliation[a]{SOLEIL Synchrotron, L’Orme des Merisiers, Départementale 128, Saint-Aubin, 91190, France}

\emailAdd{luis.manzanillas@synchrotron-soleil.fr}

\abstract{
Synchrotron radiation facilities provide highly polarized X-ray beams across a wide energy range. However, the exact type and degree of polarization varies according to the beamline and experimental setup. To accurately determine the angle and degree of linear polarization, a portable  X-ray polarimeter has been developed. 
This setup consists of a Silicon Drift Detector that rotates around a target made of high-density polyethylene. The imprint generated in the angular distribution of scattered photons at a 90-degree angle from the target has been exploited  to determine the beam polarization.  
Measurements were conducted at the GALAXIES beamline of the SOLEIL Synchrotron.
The expected angular distribution of the scattered photons for a given beam polarization was obtained through simulations using the Geant4 simulation toolkit. An excellent agreement between simulations and the collected data has been obtained, validating the setup and enabling a precise determination of the beam polarization.
}


\begin{document}
\maketitle
\flushbottom

\section{Introduction}
\label{sec:intro}

Synchrotron facilities are well known sources of highly focused X-ray beams across a wide energy range. X-ray beams are valuable tools for studying many aspects of the structure of matter at the atomic and molecular scale \cite{Mobilio2015}. X-ray beams at synchrotron facilities are characterized by their high intensity, natural narrow angular collimation, pulsed time structure and high degree of polarization.  

The radiation emitted at synchrotron radiation facilities using bending magnet, wiggler and planar undulator sources is predominantly linear polarized in the horizontal plane of emission. Other sources such as helical undulators or elliptically polarized wigglers can alter the polarisation of the emitted light and X-ray phase retarders can be used to convert horizontal linear polarisation into elliptical, circular or vertically linearly polarised light. 
In certain instances, the beam's polarization can be tailored to meet the requirements of specific experiments where certain polarization conditions are needed \cite{Beaurepaire2001}. 
In particular, this is required when the magnetic properties of materials are studied using X-ray magnetic circular dichroism or similar techniques, where the difference in absorption of left and right circularly polarized X-rays by a magnetic sample is exploited \cite{FUNK20053}. 
The accurate determination of the photon polarization from a given source is in this case of paramount importance and falls within the domain of polarimetry.
X-ray polarimetry is of great interest not only within the context of synchrotron facilities but also in other fields such as optics and astrophysics \cite{DelMonte:2023irw,Muleri_2014}.

By detecting changes in the plane and amplitude of polarization, polarimetry instruments offer a deeper understanding of the sources and processes that emit or interact with polarized light. This approach is used when  characterizing materials at atomic and molecular scales at synchrotron facilities by means of polarized X-rays. For X-rays with energies below 30~keV the dominant  photon-matter interaction process is the photo-electric effect \footnote{X-rays with energies below 30~keV predominantly interact through the photoelectric effect. Nevertheless, the emitted electrons are re-absorbed within the material and therefore these electrons can not be used for polarimetry studies.}.
However, Rayleigh and Compton scattering becomes non negligible depending on the target material at energies higher than  5~keV. This allows to develop polarimetry instruments that exploit the scattering of X-rays in a target.  
When looking for a target material candidate, high-density poly-ethylene (HD-PE) is of particular interest due to its low atomic number (Z) and non negligible Rayleigh and Compton cross sections. These two characteristics allow to achieve a low energy threshold due to the scattering/absorption probability ratio. Additionally, its homogeneous and non-crystalline structure makes it an excellent diffuser.
Around 10~keV, photons interact within the HD-PE through Compton and Rayleigh scattering in equal proportions. Below 10 keV, Rayleigh scattering is dominant, while above this energy, Compton scattering takes dominance. 
In conclusion, HD-PE is an ideal material to be used as a target to construct a Rayleigh-Compton polarimeter for X-rays with energies above 5~keV.

According to the inherent property of Compton and Rayleigh scattering, photons preferentially scatter perpendicular to the plane of linear polarization \cite{Chattopadhyay:2021mbb}. Thus, linearly polarized photons will leave an imprint with polarization information on the angular distribution of the scattered photons.

\begin{figure}[ht]
    \centering
    \includegraphics[width=0.9\textwidth]{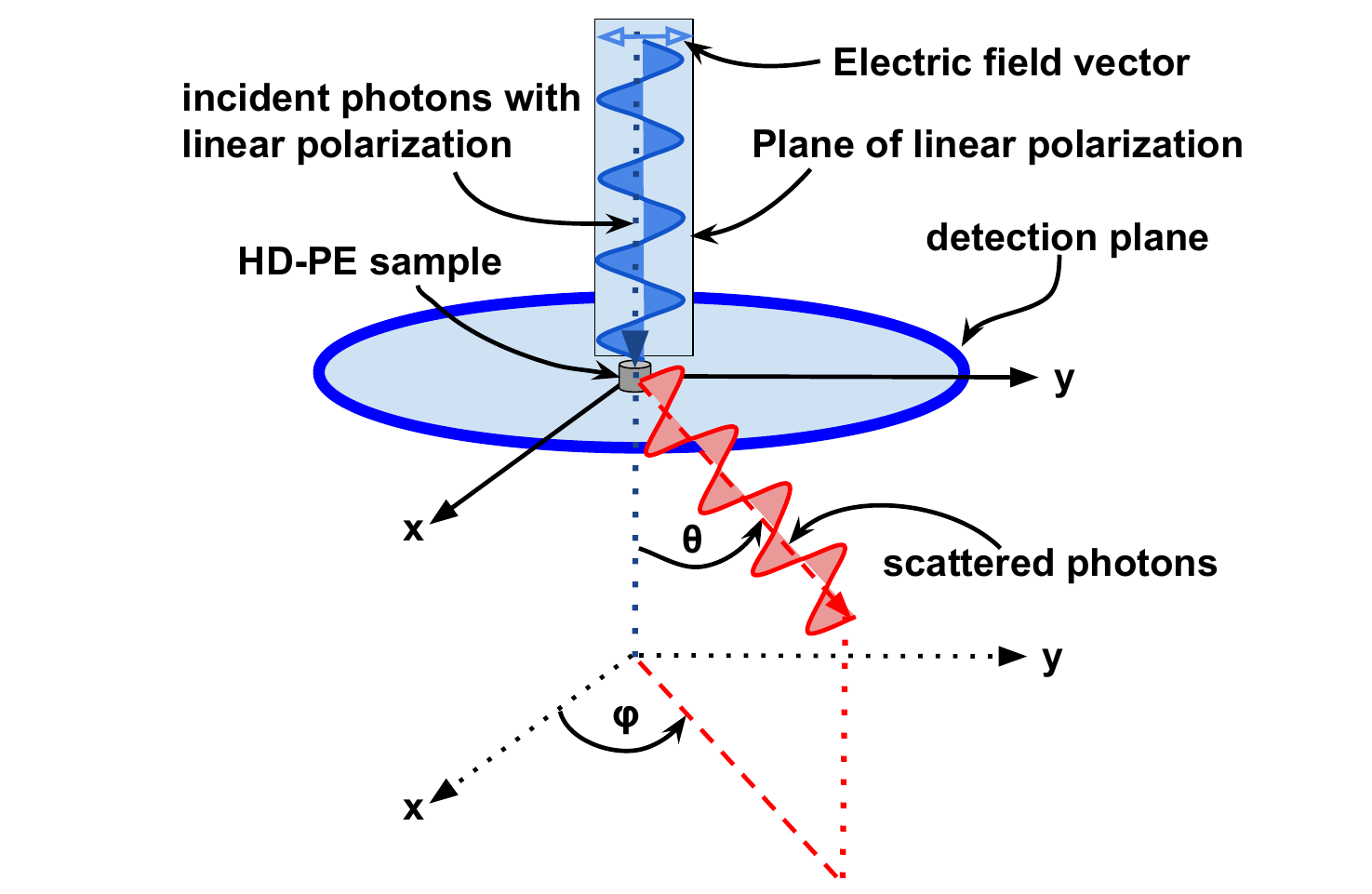}
    \caption{Sketch of scattering of linearly polarized photons in a HD-PE sample. }
    \label{fig:sketch_scattering}
\end{figure}

The angular distribution of scattered photons with an angle $\theta=90^{\circ} $ as shown in Figure~\ref{fig:sketch_scattering} is given by 

\begin{equation}
N(\varphi)=A + B \cos^2{(\varphi -\varphi_0)}
\label{eq:Events_phi}
\end{equation}
with $\varphi$ being the angle with respect to the $x$-axis,
$\varphi_0 \pm\pi/2$ defining the plane of linear polarization,  $A$ and $B$ being a constant and the amplitude of the angular distribution, respectively. The modulation parameter $M$ for a given degree of linear polarization is defined using the maximum and minimum of the angular distribution of events as

\begin{equation}
M=\frac{N_{Max}-N_{Min}}{N_{Max}+N_{Min}}=\frac{B}{2A+B}
\label{eq:modulation}
\end{equation}
In consequence, the development of instruments capable of accurately detecting the angular distribution of events scattered in a target material allows one to assess the plane and degree of linearly polarized photons.

In this paper, we outline the development of a compact Rayleigh-Compton polarimeter designed for hard X-rays. In Section~\ref{sec:setup}, we describe the polarimeter setup and the simulations conducted using the Geant4 simulation toolkit to anticipate the modulation pattern for a given degree of linear polarization. Moving on to Section~\ref{sec:galaxies_measurements}, a brief overview of the GALAXIES beamline and the data acquisition protocols are provided. The outcomes from the initial measurement campaign at  GALAXIES in 2023 are detailed in Section~\ref{sec:results}. Finally, the concluding remarks of our study can be found in Section~\ref{sec:conclusions}.

\section{Experimental setup}\label{sec:setup}

The device developed in this work was inspired from a polarimeter built by other authors using two Cadium-Telluride detectors and a fixed diffuser~\cite{TOKANAI2004446}.
In this work we have developed a new polarimeter consisting of a Silicon Drift Detector (SDD) (Model XR-100SDD-FAST from AMPTEK) that rotates around a sample made of HD-PE which acts as a diffuser medium. The HD-PE sample is mounted on a 360-degrees rotation  stage to average sample anisotropies during measurements.
The sensitive area of the SDD detector is 25~mm${^2}$, and its thickness is 500~$\mu m$. The sensitive area of the detector is placed perpendicular to the direction of the beam at 62~mm from the center of the HD-PE sample.  
This HD-PE sample has a cylindrical geometry with a diameter of 3~mm and length of 20~mm.
The detector and HD-PE sample are attached to the setup by means of 3D printed supports.
\begin{figure}[ht]
    \centering
    \includegraphics[width=\textwidth]{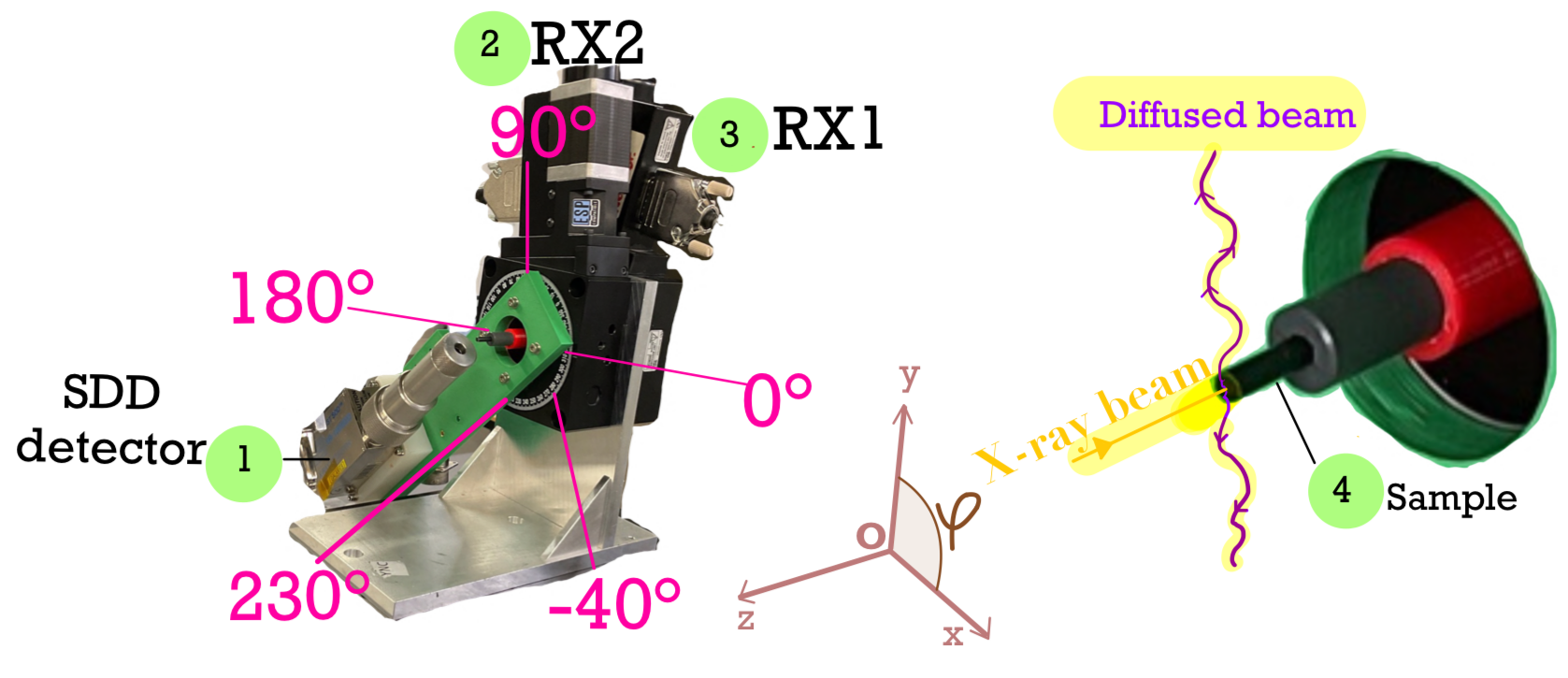}
    \caption{Left: Polarimeter setup developed at SOLEIL. The sensitive area of the SDD detector (1) is placed at 62 mm from the HD-PE sample (4). Motor RX2 (2) is used to rotate the detector around the sample, while motor RX1 (3) is used to rotate the sample. The angular resolution of the motors is 15 mili-degrees. Because of the cabling and supports, the detector can only cover an angular section of 270 degrees. Right: Zoom into the HD-PE sample (4) and its 3D printed support. }
    \label{fig:setup}
\end{figure} 

Figure~\ref{fig:setup} shows the polarimeter setup.
Due to its mounting, cabling and supporting arm, the detector has a limited range of motion of 270 degrees. 
Despite this limitation, this range adequately covers at least a maximum and minimum of the anticipated angular distribution. 
Consequently, it furnishes enough data to determine both the plane and degree of linear polarization.

The SDD output is connected to an Xspress3M Digital Pulse Processor (DPP)~\cite{QuantumXspress3,FARROW1995567}, which digitizes X-ray signals and provides the input count rate (ICR) and output count rate (OCR) of events, and also create an energy spectrum from signal amplitudes. The Xspress3M DPP has been calibrated with an $^{55}$Fe X-ray source (main X-ray line at 5.9~keV) to get a spectrum binning of 10~eV and an energy resolution of 127~eV (FWHM) at 5.9~keV at an Input Count Rate of 1-10~kcps. The Xspress3M DPP and the two rotation motors (called RX1 for the HD-PE sample and RX2 for the SDD detector) are integrated into the SOLEIL TANGO control system~\cite{TANGO1999,Rubio-Manrique:2015evh}. During a measurement, the motor RX2 scans a wide range of angles (270 degrees) with a step of 10 degrees. At each RX2 angle, the motor RX1 rotates the sample and the Xspress3M acquires data during 10 seconds. Then, data (RX1 angle, ICR, OCR and energy spectrum) are recorded in a file.

\subsection{Simulation of the setup}
\label{sec:simsetup}
\begin{figure}[ht]
    \centering
    \includegraphics[width=\textwidth]{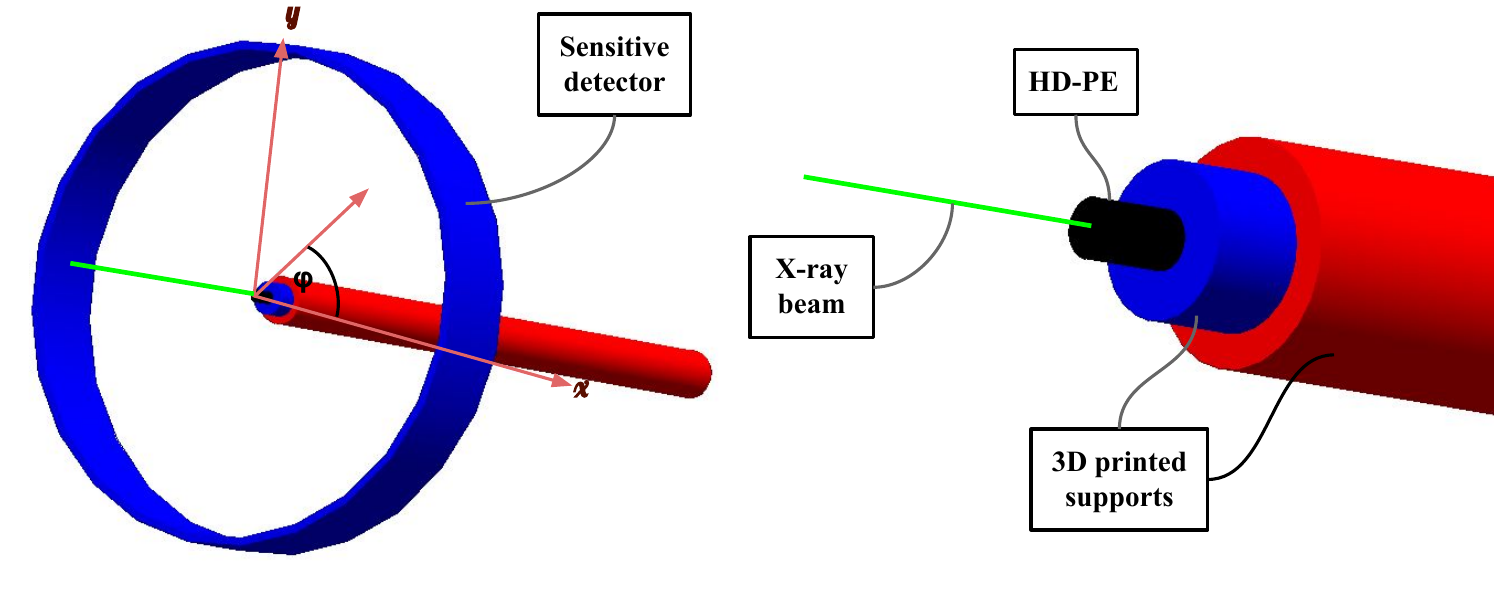}
    \caption{Sketch of Geant4 implementation of the polarimeter setup}
    \label{fig:setup_geant4}
\end{figure}

In order to predict the angular distribution for a given degree of linear polarization, the Geant4 simulation toolkit \cite{GEANT4:2002zbu} employing its dedicated low energy physics list \textit{G4EmLivermorePolarizedPhysics}  was used. 
A detailed HD-PE sample with its 3D printed supports was implemented in Geant4. To simplify data treatment, as well as to maximize CPU performance, a circular detector around the sample as shown in Figure \ref{fig:setup_geant4} was implemented.  The expected angular distribution for a given degree of linear polarization is obtained by analyzing the interactions of photons in the sensitive detector.

The polarization of particles in Geant4 is included using the Stokes formalism.
Assuming the particle reference frame, i.e. with the momentum of the particle pointing along the z direction, the stokes vector $S$ can be parameterized as follows
\begin{equation} \label{eq_stokes}
\begin{split}
S_1 & = \cos(2\psi)\cos(2\chi) \\
S_2 & = \sin(2\psi)\cos(2\chi) \\ 
S_3 & = \sin(2\chi)
\end{split}
\end{equation}
with $\psi$ and $\chi$ being the angles of the Poincare sphere.
The factors of two before $\psi$ and $\chi$ correspond, respectively, to the fact that any polarization ellipse is indistinguishable from one rotated by 180 degrees.
From equation \ref{eq_stokes} one can obtain the particular case of linear polarization when $\chi=\pi$ 
\begin{equation} \label{eq_stokes_linear}
\begin{split}
S^L_1 & = \cos(2\psi) \\
S^L_2 & = \sin(2\psi) \\ 
S^L_3 & = 0
\end{split}
\end{equation}

$\psi$ defining the plane of linear polarization. Then, with $\psi = \pi$   or $\psi=\pi/2$ one obtains horizontal ($S=[1,0,0]$) and vertical ($S=[0,1,0]$) polarization, respectively. On the other hand, circular polarization is obtained when $\chi=\pi/2$ ($S=[0,0,\pm1]$). We decompose the polarization of the beam in linear and elliptical or circular polarization components
\begin{equation} \label{eq_stokes_amount_polarization}
S = \mathbb{P}S^L + (1-\mathbb{P})S^\epsilon
\end{equation}
with $\mathbb{P}$ being the degree of linear polarization. 

\begin{figure}[ht]
    \centering
    \includegraphics[width=0.47\textwidth]{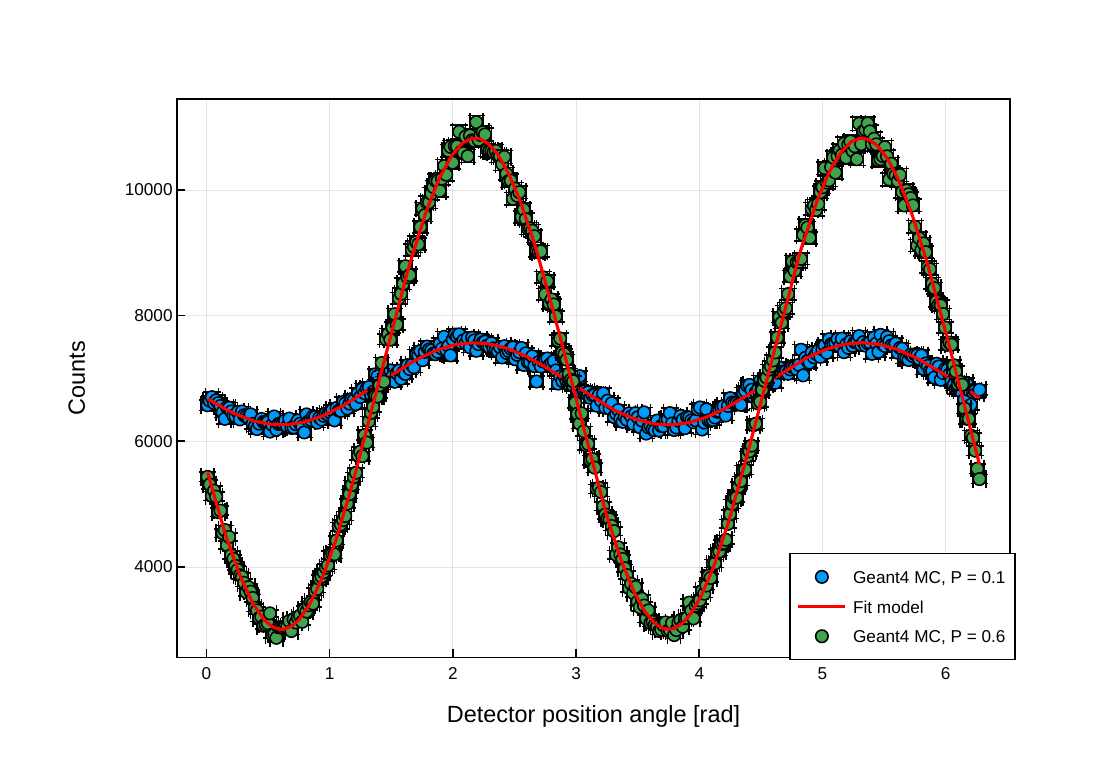}
    \includegraphics[width=0.5\textwidth]{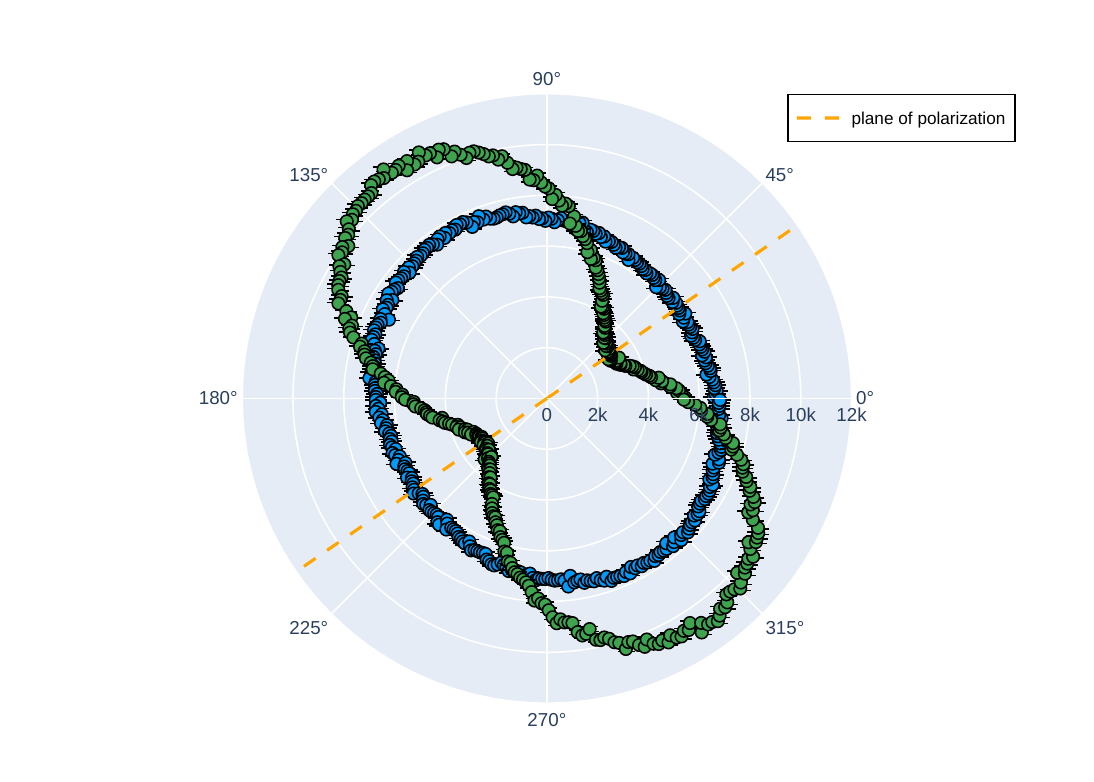}
    \caption{Left: Simulated angular distribution generated by 7 keV photons assuming a degree of linear polarization $\mathbb{P}$ of 10\% (blue) and 60\% (green), respectively. Right: Same distributions in polar coordinates. The plane of oscillation of the electric field for the fraction of linearly polarized photons is indicated with the dashed line in orange.}
    \label{fig:mc_polarization_examples_geant4}
\end{figure}

\begin{figure}[ht]
    \centering
    \includegraphics[width=0.7\textwidth]{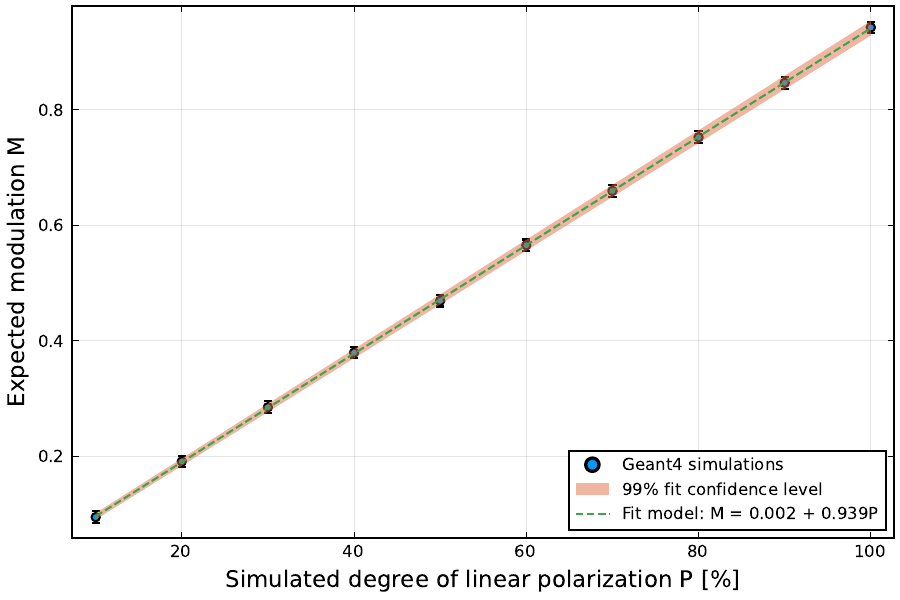}
    \caption{Geant4 predicted modulation as function of degree of linear polarization for photons of 7~keV. A linear dependence is observed between $M$ and $\mathbb{P}$.}
    \label{fig:mc_polarization_modulation_geant4}
\end{figure}

Figure~\ref{fig:mc_polarization_examples_geant4} shows the expected angular distributions for two degrees of linear polarization $\mathbb{P}$ and the plane of linear polarization (35\textdegree,215\textdegree). These distributions can be  fitted using equation \ref{eq:modulation} as shown in Figure~\ref{fig:mc_polarization_examples_geant4}~(Left) in order to extract the expected modulation $M$ for a given degree of linear polarization. 
The dependence of $M$ with $\mathbb{P}$ is shown in Figure~\ref{fig:mc_polarization_modulation_geant4}. A linear relation is observed, which was fitted using a linear model as follows
\begin{equation} \label{eq_linear_fit}
M(\mathbb{P}) = a + b \mathbb{P}
\end{equation}
with $a$ and $b$ being parameters that are dependent on the setup and beam energy.

\section{Polarization measurement campaign at the GALAXIES beamline}
\label{sec:galaxies_measurements}

\subsection{Beamline description}
GALAXIES is an in-vacuum undulator hard X-ray micro-focused beamline dedicated to studying the electronic structure of materials with high energy resolution at  the SOLEIL synchrotron \cite{Rueff:ie5123,Ablett:il5013,CEOLIN2013188,JPRUEFFHAXPES}. It is dedicated to both hard X-ray photo-electron spectroscopy ((HAXPES)) and inelastic X-ray scattering techniques under both non-resonant (NR-IXS) and resonant (RIXS) conditions.
The beamline is optimized to operate in the 2.3 - 12 keV energy range with high resolution. 
The full-width-half-maximum beam size on the target is about $100 \text{ [horizontal]}\times 30\text{ [vertical]}~\mu\text{m}^2$.

At GALAXIES the beam polarization can be manipulated by means of a diamond X-ray phase retarder (XPR) in the Laue asymmetric transmission geometry. The GALAXIES XPR consist of a 500 $\mu m$-thick single-crystal synthetic diamond (111). XPRs exploits the birefringence of perfect crystals under the condition of Bragg diffraction \cite{authier_bragg_difraction,BELYAKOV1989526}.
By altering the angle of incidence of the beam around the Bragg angle, circular and different states of linear polarization can be generated.

\subsection{Experimental procedure}
\begin{figure}[ht]
    \centering
    \includegraphics[width=0.47\textwidth]{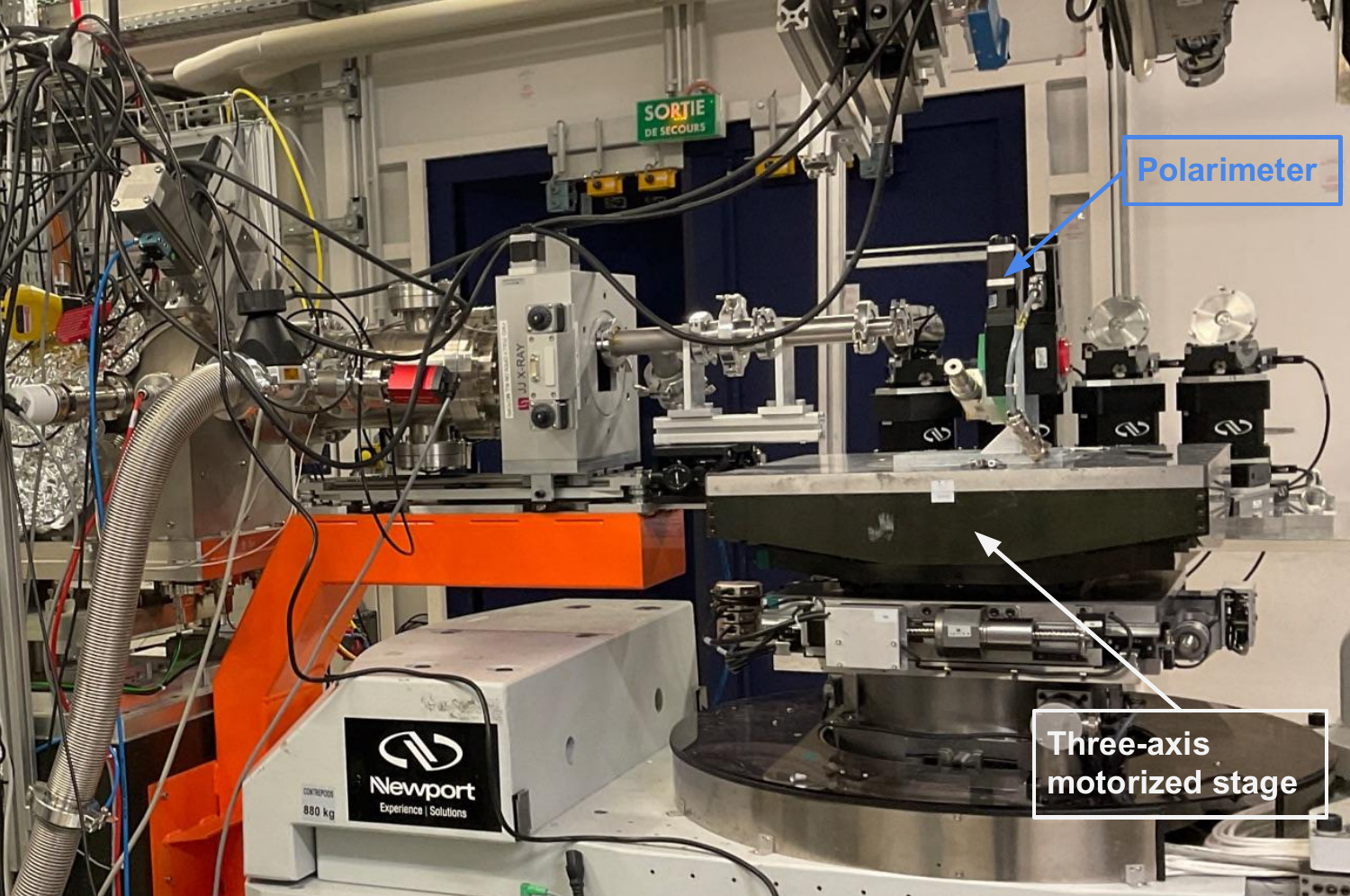}
    \includegraphics[width=0.47\textwidth]{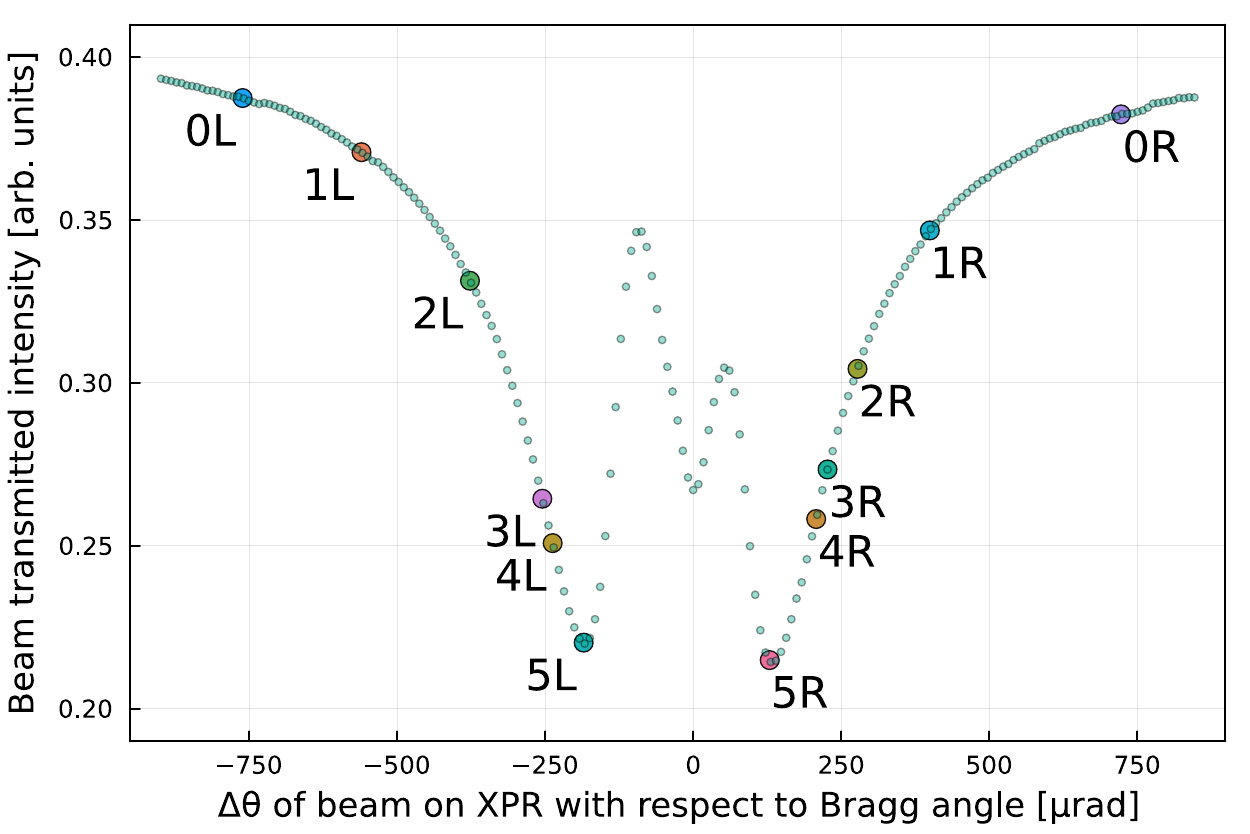}
    \caption{Left: Polarimeter setup mounted on a three-axis motorized stage in the GALAXIES RIXS experimental hutch. Right: XPR profile obtained by recording the vertically scattered X-ray radiation as a function of XPR angle (see text for more details).}
    \label{fig:qwp_points_galaxies}
\end{figure}

The experimental setup was installed on a three-axis motorized stage in the GALAXIES RIXS experimental hutch \cite{Ablett:il5013} as shown in Figure~\ref{fig:qwp_points_galaxies}~(Left), capable of sub-micrometric precision movement.
Prior to commencing data collection for polarimetry studies, two tasks must be completed: First, the Bragg angle needs to be determined, and second, the beam must be aligned to hit the center of the sample, as detailed in the subsequent subsection. At GALAXIES, the vertically scattered X-ray radiation from a thin Kapton foil, placed immediately downstream of the XPR, is recorded by a photodiode orientated in the vertical direction in order to obtain a profile like that shown in Figure~\ref{fig:qwp_points_galaxies}~(Right). From this profile, and the incident X-ray energy, the XPR Bragg angle can be determined.

In the curve shown Figure~\ref{fig:qwp_points_galaxies}~(Right), horizontal polarization dominance is expected on the left (point 0L) and right (point 0R) sides, and vertical linear polarization dominance at the two minima (points 5L and 5~R) \cite{freeland2002,authier_bragg_difraction,Hirano_1997,GILES1994622,Francoual_2013}. Furthermore, circular polarization is expected to maximize approximately at the halfway point of the maximum, going from one circular polarization handiness to the opposite, on the left and right side, respectively.

Finally, data were collected  for 10 polarization states by selecting 10 XPR angles. These angles were chosen from both sides of the beam-transmitted intensity curve, as shown in Figure~\ref{fig:qwp_points_galaxies}~(Right). For each point a scan was taken by changing the position of the SDD detector in the range from -40~degrees to 230~degrees with steps of 10~degrees  and 10~s of data acquisition in each point.

\subsection{Beam alignment}

The second step prior to determining the angular distribution of events involves aligning the beam to accurately target the center of the HD-PE sample. Initially, a coarse alignment method is utilized. Following this, the sample is temporarily removed, and a thin YAG:Ce X-ray scintillator is attached to cover the designated sample area. Subsequently, scans are executed along the X and Z axes using the motors within the setup station. The center of the sample is identified utilizing a camera that detects the light emitted by the scintillator. 

\begin{figure}[ht]
    \centering
    \includegraphics[width=0.6\textwidth]{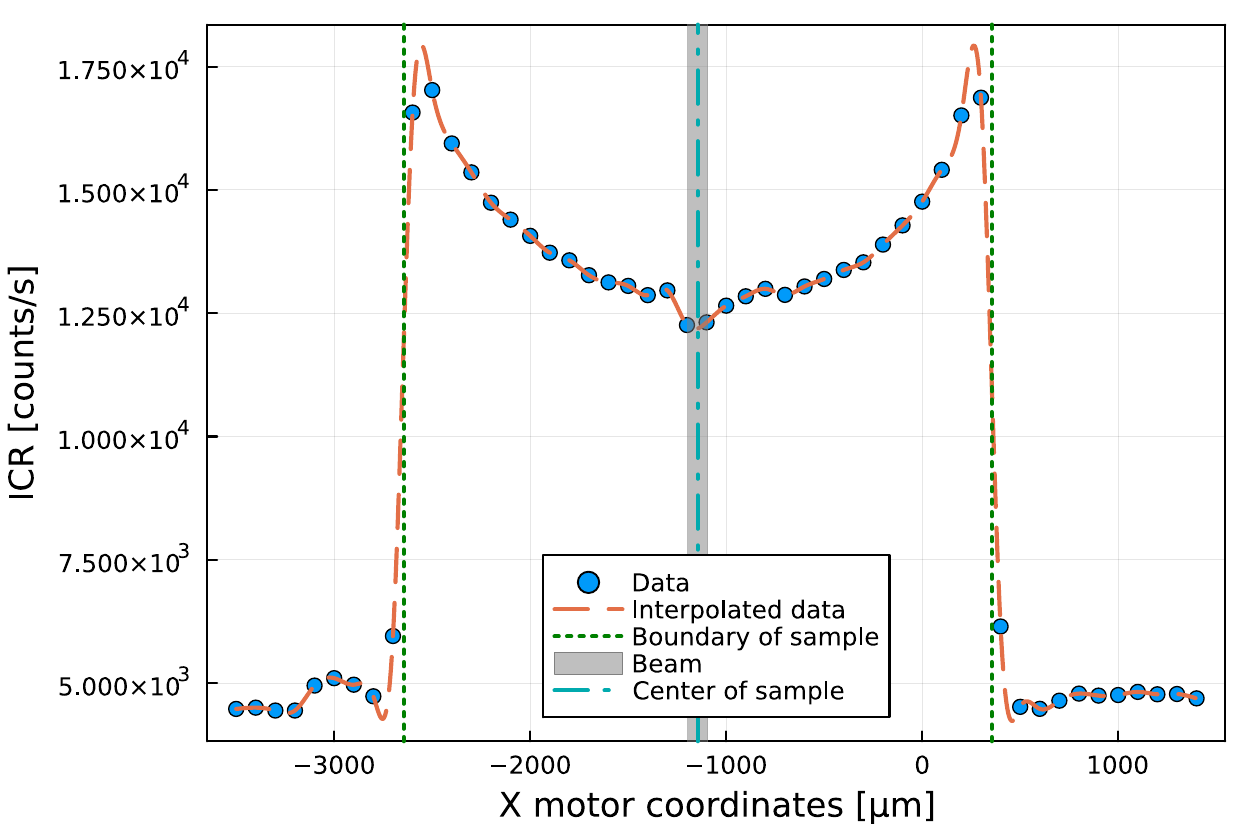}
    \caption{Number of detected photons as a function of the X motor coordinates.}
    \label{fig:qwp_points_align}
\end{figure}

A more refined alignment is then achieved by repositioning the sample within the setup and conducting scans in both horizontal and vertical directions using steps of $100~\mu m$. During the horizontal scan, the detector is oriented at 90 degrees with respect to the horizontal axis, while for the vertical scan, it is placed at 0 degrees relative to the vertical axis. In both instances, a plateau is observed when plotting the event count rate against the position, as illustrated in Figure~\ref{fig:qwp_points_align}. Finally, the center of the sample is determined through data interpolation and by identifying the sample's boundaries, which are determined by identifying the points corresponding to half of the maximum count rate.

The shape of the plateau observed in Figure~\ref{fig:qwp_points_align} illustrates the influence of 7~keV photon attenuation within the HD-PE sample. A difference of approximately 32\% is noticeable when the beam impacts the center of the sample compared to when it strikes the border of the sample. Consequently, any misalignment of the beam could potentially give rise to artifacts and introduce asymmetries into the angular distribution of events. This effect is anticipated to become more pronounced at lower energies and diminish at higher energies.

\section{Analysis and results}
\label{sec:results} 
\begin{figure}[ht]
    \centering
    \includegraphics[trim={2cm 0 0 0},clip,width=\textwidth]{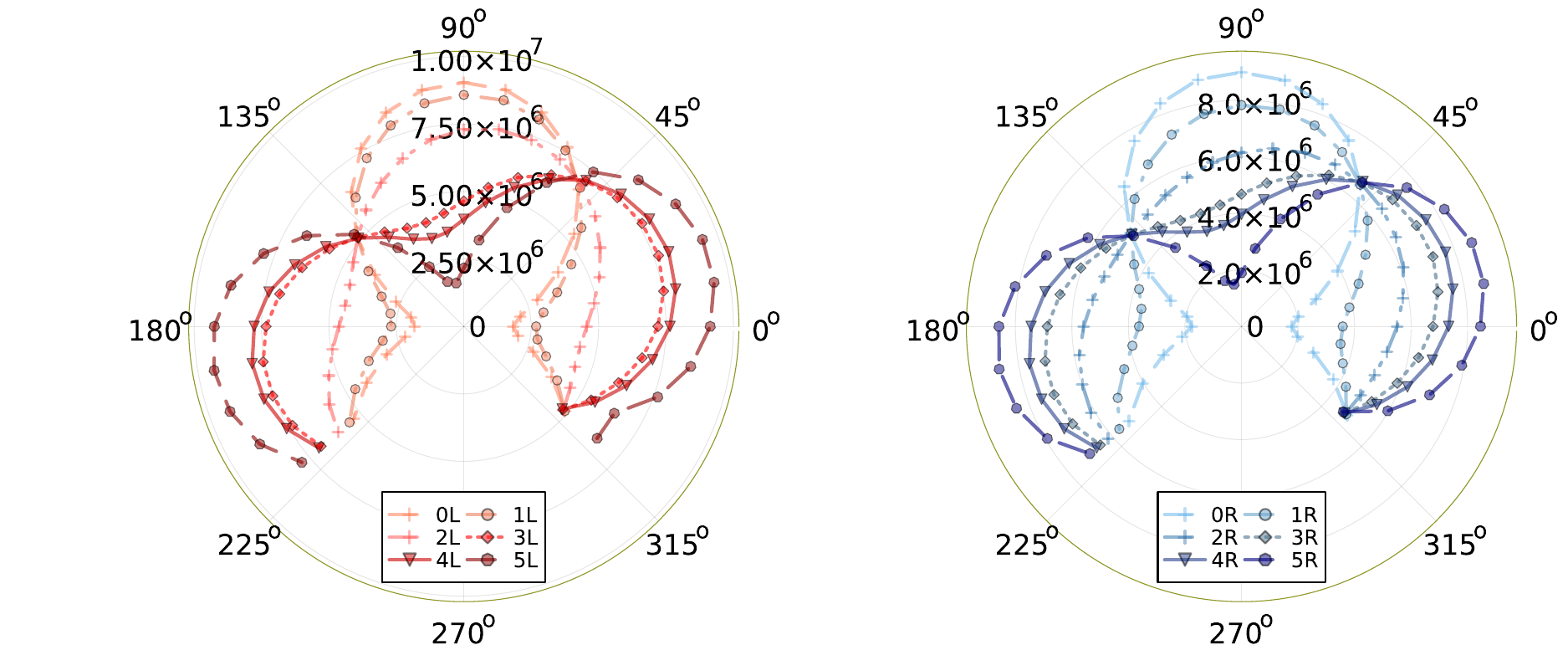}
    \caption{Number of detected photons as function of the angular position of detector for all the positions shown in Figure~\ref{fig:qwp_points_align}}
    \label{fig:distribution_all_qwp_points}
\end{figure}

The angular distribution of the ICR for each polarization state selected in Figure~\ref{fig:qwp_points_galaxies}~(Right) is depicted in Figure~\ref{fig:distribution_all_qwp_points}. As anticipated, the points positioned at the far right (0R) and far left (0L) of the curve shown in Figure~\ref{fig:qwp_points_galaxies}~(Right) exhibit a behavior consistent with horizontal linear polarization, whereas the points situated within both minima display vertical linear polarization. Conversely, the degree of linear polarization diminishes for points situated between the minima and the far left or far right extremes. 

\begin{figure}[ht]
    \centering
    \includegraphics[width=0.47\textwidth]{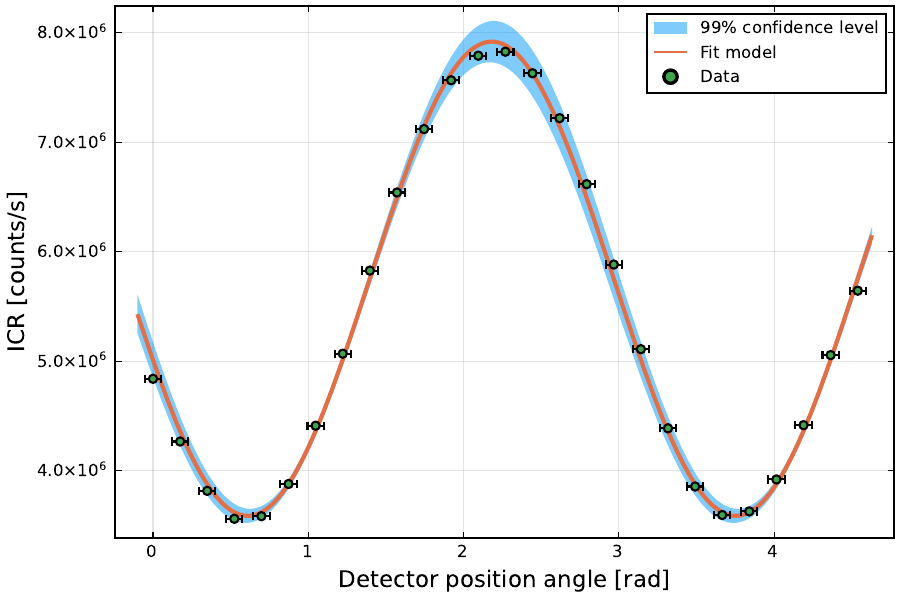}
    \includegraphics[width=0.47\textwidth]{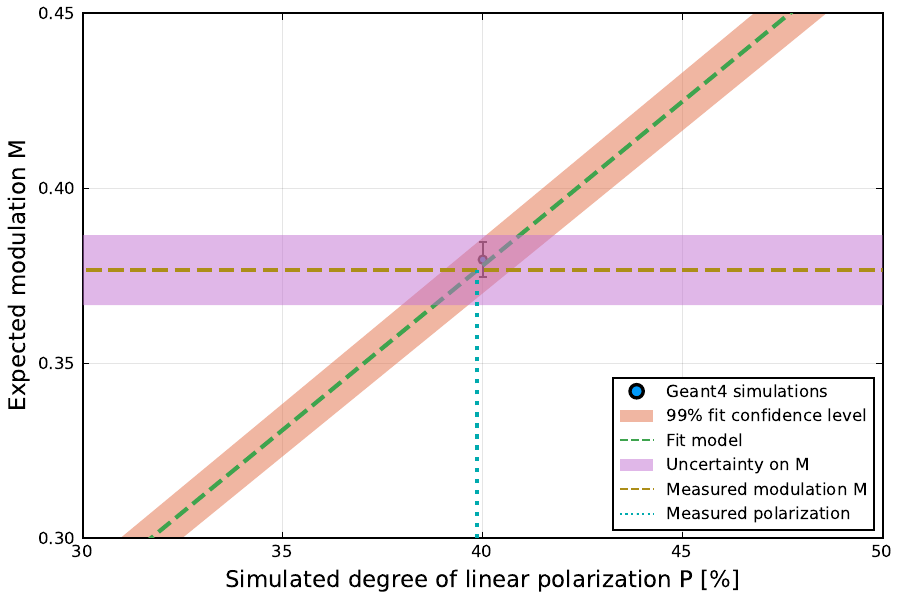}
    \caption{Left: Measured angular distribution of the detected number of events for a state of polarization corresponding to point 1R of Figure~\ref{fig:qwp_points_galaxies}~(Right). The distribution was fitted with equation \ref{eq:modulation} as shown with the red line. Right: Measured modulation parameter $M$ and it's intersection with the predicted $M-\mathbb{P}$ curve obtained from simulations to determine the degree of linear polarization. The uncertainty associated with the measurement of $M$ is represented by the pink horizontal band, while the uncertainty related to the predicted $M-\mathbb{P}$ curve is indicated by the orange band.}
    \label{fig:data_fit}
\end{figure}

The angular distributions of the ICR shown in Figure~\ref{fig:distribution_all_qwp_points} have been employed to derive both the degree of linear polarization $\mathbb{P}$ and the plane of linear polarization. For this purpose, each distribution was fitted using equation~\ref{eq:modulation}. Subsequently, the modulation parameter $M$ and the plane of linear polarization ($\varphi_0\pm\pi/2$) were determined for each respective case.
Figure~\ref{fig:data_fit}~(Left) shows an example of a data set (position 1R in Figure~\ref{fig:qwp_points_galaxies}~(Right)) fitted with equation~\ref{eq:modulation}. The modulation parameter $M$ was derived from the values of $A$ and $B$ obtained from the fit. After measuring $M$, it can be juxtaposed with the predictions generated by Geant4 to obtain the degree of linear polarization $\mathbb{P}$, as depicted in Figure~\ref{fig:data_fit}~(Right). 

\begin{figure}[ht]
    \centering
    \includegraphics[width=0.47\textwidth]{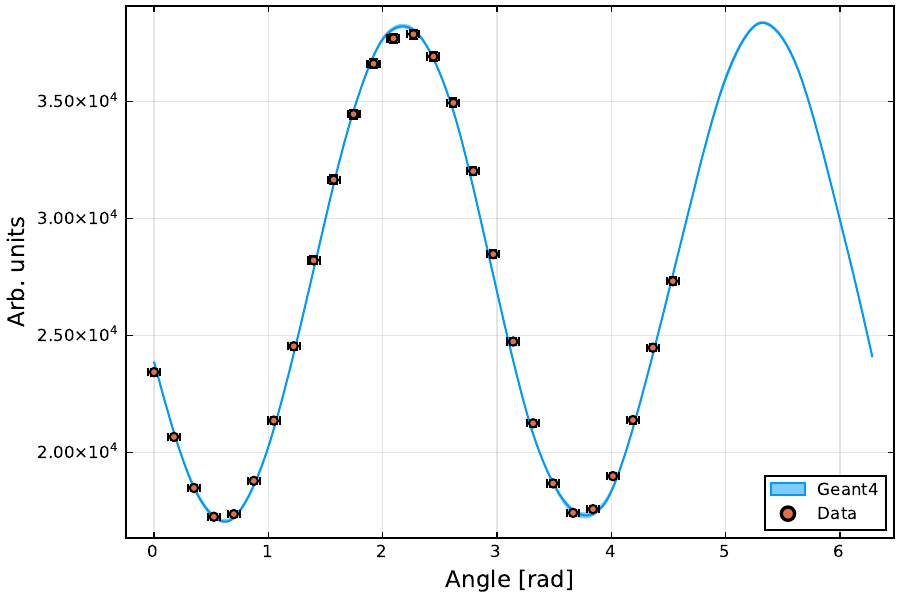}
    \includegraphics[width=0.47\textwidth]{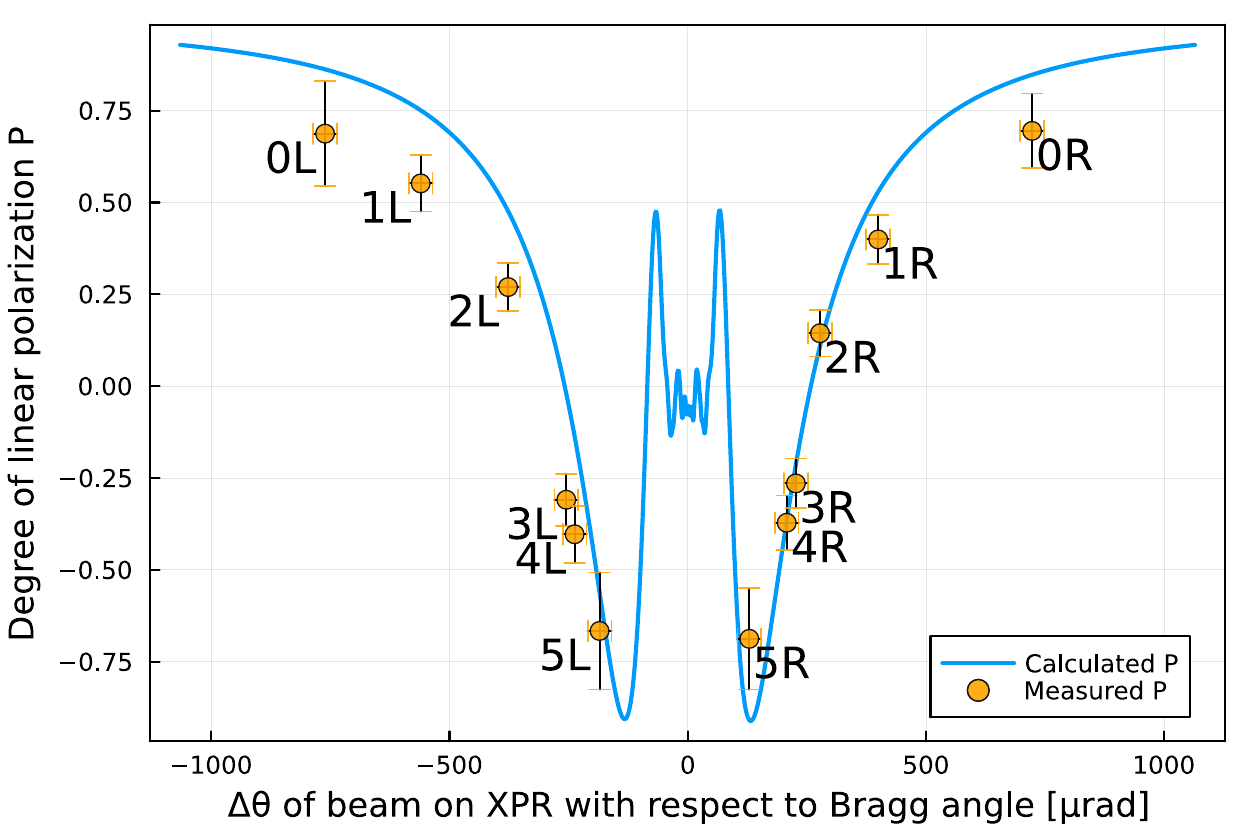}
    \caption{Left: Data - MC comparison for dataset of point 1R. Right: Measured degree of linear polarization compared with prediction. The sign of P indicate horizontal (positive) or vertical (negative) polarization dominance.
    A systematic error of 5\% is included in the error bar}
    \label{fig:mc_data_compare}
\end{figure}

The measured plane of linear polarization, along with the derived degree of linear polarization $\mathbb{P}$, can serve as input for the simulations to generate a modulation pattern and compare it with the data.
As shown in Figure~\ref{fig:mc_data_compare}~(Left), a comparison is presented between the data and the Geant4 predictions for point 1R. An excellent agreement was achieved, thereby validating our methodology.

Figure~\ref{fig:mc_data_compare}~(Right) shows the calculated degree of linear polarization for photons of 7~keV selected by the XPR and the measured values reported in this work. An excellent agreement is found at small $\Delta\theta$ with respect to the Bragg angle.  At larger $\Delta\theta$, the agreement is less favorable, but the measurements still follow the expected trend. These differences can be attributed to XPR repeatability issues stemming from mechanical problems that were identified after the completion of this work and have subsequently been rectified. The measurement campaign reported in this work helped to identify and address these mechanical problems.

The results show that the point with the lowest degree of linear polarization corresponds to point 2R. Around this point, circular polarization is expected. The highest degree of linear polarization is observed at both minima (OL, OR) and the two extreme points (5L, 5R), reaching a degree of linear polarization of approximately 70\%.
Most of the errors have been estimated to be below the 2\% level, though they are higher ($\sim10\%$) in certain cases due to missing information in the lower part of the angular distribution.
A modification of the setup that would enable complete coverage of 360 degrees is currently under investigation and will be implemented in an upgraded version of the setup.
The values of the results of degree of linear polarization shown in Figure~\ref{fig:mc_data_compare}~(Right) together with the plane of linear polarization for each measurement are summarized in Table~\ref{table:1}.

\begin{table}[ht]
\centering
\begin{tabular}{ |c|c|c|c| } 
 \hline
 Beam-XPR angle & Modulation $M$ & Plane of polarization [\textdegree] & Degree of linear polarization $\mathbb{P}$ [\%] \\ [0.5ex] 
 \hline\hline
0L	&	0.65	$\pm$	0.04	&	(	1	,	181	)	$\pm$	1	&	68.8	$\pm$	4.3 	\\	\hline
1L	&	0.52 $\pm$	0.02	&	(	2.8	,	182.8	)	$\pm$	0.8	&	55.3	$\pm$	1.8	 	\\	\hline
2L	&	0.25	$\pm$	0.01	&	(	12.5	,	192.5	)	$\pm$	1.3	&	27.1	$\pm$	0.7 		\\	\hline
3L	&	0.29	$\pm$	0.01	&	(	68.8	,	248.8	)	$\pm$	2	&	31	$\pm$	1.2	\\	\hline
4L	&	0.38	$\pm$	0.02	&	(	72.6	,	252.6	)	$\pm$	1.8	&	40.4	$\pm$	1.9	\\	\hline
5L	&	0.63	$\pm$	0.09	&	(	78.3	,	258.3	)	$\pm$	2.6	&	66.7	$\pm$	10.1	\\	\hline
5R	&	0.65	$\pm$	0.07	&	(	78.9	,	258.9	)	$\pm$	1.7	&	68.9	$\pm$	7.9	\\	\hline
4R	&	0.35	$\pm$	0.01	&	(	73.6	,	253.6	)	$\pm$	1.7	&	37.3	$\pm$	1.5	\\	\hline
3R	&	0.25	$\pm$	0.01	&	(	68.5	,	248.5	)	$\pm$	1.9	&	26.6	$\pm$	0.9	\\	\hline
2R	&	0.14	$\pm$	0.01	&	(	33.1	,	213.1	)	$\pm$	2.9	&	14.5	$\pm$	0.5	\\	\hline
1R	&	0.38	$\pm$	0.01	&	(	5	,	185	)	$\pm$	0.8	&	40.1	$\pm$	0.8	\\	\hline
0R	&	0.65	$\pm$	0.04	&	(	0.7	,	180.7	)	$\pm$	1	&	69.6	$\pm$	4.3	\\	\hline
 \hline
\end{tabular}
\caption{Summary of polarimetry results.}
\label{table:1}
\end{table}

\section{Conclusions}
\label{sec:conclusions} 
A polarimetry setup has been developed at  SOLEIL. This polarimeter exploits the Rayleigh-Compton scattering of X-rays at a 90-degree angle from a high-density polyethylene sample. The resultant scattered photons are detected by an Silicon Drift Detector, which rotates around the high-density polyethylene sample.
The distribution of scattered photons, as detected by the Silicon Drift Detector, was used to measure the modulation parameter $M$.

A measurement campaign was conducted at GALAXIES, where the degree and plane of linear polarization can be adjusted using an X-ray phase retarder. Through the utilization of the Geant4 simulation toolkit, the modulation parameter $M$ and its correlation with the degree of linear polarization $\mathbb{P}$ was estimated. By comparing the measured modulation parameter $M$ with the predicted values, the degree of linear polarization was determined. The comparison between the data and the Monte-Carlo simulations demonstrates an excellent agreement, thus validating our methodology and setup.

\acknowledgments
The authors would like to express their gratitude to C. Menneglier and B. Kanoute from the detector group of SOLEIL for their invaluable assistance in the production of mechanical supports and cabling and to A. Paquis for his contributions preparing a first version of the setup.


\bibliographystyle{JHEP}
\bibliography{biblio.bib}

\end{document}